\documentclass[aps,prl,nofootinbib,superscriptaddress,twocolumn]{revtex4}
\usepackage{amssymb,amsmath,latexsym,graphics, graphicx,epsfig} 
\usepackage{multirow}
\usepackage{slashed}

\newcommand{\gsim}{\lower.7ex\hbox{$\;\stackrel{\textstyle>}{\sim}\;$}}
\newcommand{\lsim}{\lower.7ex\hbox{$\;\stackrel{\textstyle<}{\sim}\;$}}

\def\LL{{\cal L}}
\def\OO{{\cal O}}

\def\eg{{\it e.g.~}}

\newcommand{\TeV}{\,\mathrm{TeV}}
\newcommand{\GeV}{\,\mathrm{GeV}}

\newcommand{\bef}{\begin{figure}[htbp]\begin{center}}
\newcommand{\eef}{\end{center}\end{figure}}

\newcommand{\bea}{\begin{eqnarray}}
\newcommand{\eea}{\end{eqnarray}}

\newcommand{\mzt}{m_{Z^0}}

\begin{document}

\title{
\begin{flushright}
\mbox{
\small
SLAC-PUB-14131}
\end{flushright}
Model Independent Bounds on Kinetic Mixing	}
\author{Anson Hook, Eder Izaguirre, Jay G. Wacker}
\affiliation{
Theory Group, SLAC,  Menlo Park, CA 94025
}
\begin{abstract}
New Abelian vector bosons can kinetically mix with the hypercharge gauge boson of the Standard Model.  This letter computes
the model independent limits on vector bosons with masses from $1 \GeV$ to $1\TeV$.
The limits arise from the numerous $e^+e^-$ experiments that have been performed in this energy range and bound the kinetic mixing by $\epsilon \lsim 0.03$ for most of the mass range studied, regardless of any additional interactions that the new vector boson may have.
\end{abstract}
\pacs{} \maketitle


The Standard Model (SM) successfully describes all known interactions of SM fermions and gauge bosons; however, there are several phenomena that motivate physics beyond the SM.  
Chief among these open questions is the identity of dark matter and its interactions with the SM.  Recent anomalies in cosmic ray and direct detection experiments have motivated the exploration of new gauge interactions in a putative dark sector
  \cite{ArkaniHamed:2008qn,Pospelov:2008jd}.
New Abelian vector bosons provide one of the most robust portals for dark matter -- SM interactions.
The new vector boson can interact with the SM, even if no SM fermions are directly charged under the additional gauge symmetry.  This interaction occurs via mixed kinetic terms between the SM's hypercharge field strength and the new Abelian field strength \cite{Holdom:1985ag}. 

The Lagrangian for a kinetically mixed $U(1)$ theory is
\bea
\label{Eq:epsilon}
\mathcal{L} = \LL_{\text{SM}} -\frac{1}{4} F_{\mu\nu}^{\prime 2} - \frac{ \sin \epsilon }{2}F_{\mu \nu}' B^{\mu \nu}  + \frac{ m^2_{A'}}{2} A_\mu^{\prime 2} + \tilde{g} J_{A'}^\mu A'_\mu 
\eea
where $F_{\mu\nu}'$ is the field strength for the new vector boson,  $B^{\mu\nu}$ is the field strength for the SM hypercharge, and $J_{A'}^\mu$ encapsulates the interactions of the $A'$ with fields in the dark sector.  The mixed kinetic term is interesting for several reasons.  First, it is a dimension 4 operator, meaning that it can be generated at high  energies without decoupling.    Second, this coupling allows communication with a secluded sector that otherwise has no interactions with SM fields.     
 These considerations have motivated a dedicated program to search for kinetically mixed vector bosons \cite{Essig:2010xa, Bjorken:2009mm, Essig:2009nc, Reece:2009un}.   

Most of the current program for discovering a kinetically mixed vector boson involves producing the state and searching for its subsequent decays. This method is promising, but has the drawback that it assumes that the searches can recognize the decay products of the $A'$.    If the dark sector has states lighter than the $A'$, then the $A'$ will preferentially decay to the dark sector over SM states because the kinetic mixing parameter almost always satisfies $\epsilon \ll \tilde{g}$.    Searching for the $A'$ by looking for the dark sector final states requires a wide-ranging search program because the dark sector may decay back to the SM in a variety of different ways, \eg lepton jets \cite{Strassler:2006im,ArkaniHamed:2008qp,Baumgart:2009tn,Alves:2009nf,Cheung:2009su}. 
Model independent searches are possible using completely inclusive searches, \eg $e^+e^- \rightarrow \gamma + X$, but these are challenging and few of these searches have actually been performed.

At low masses, the best model independent bounds arise from the $(g-2)$ measurements of the electron and muon \cite{Pospelov:2008zw}; however,  the power of $(g-2)_\mu$ begins to weaken for $ m_{A'} \gsim m_\mu$. 
    At masses far above collider energies, the $A'$ can be integrated out and its effects can be encapsulated in higher dimension operators, most importantly $S$ and $T$ \cite{Peskin:1991sw,Altarelli:1990zd}.   $e^+e^-$ colliders have probed up to $\sqrt{s}=207 \GeV$ and therefore,  the effects of the $A'$ cannot be parameterized as local higher dimension operators for masses less than this energy scale.   

This letter computes the model independent constraints on the kinetic mixing parameter, $\epsilon$, for masses between $1\GeV$ and $1\TeV$ by looking for the effects of virtual $A'$s on precision SM observables.
This approach has the benefit of not requiring any knowledge of the decay modes of the $A'$ and sets an upper limit on $\epsilon$ regardless of the behavior of the decay modes to the dark sector.
    

\section{Kinetic Mixing}
\label{Sec: Kinetic Mixing}


Kinetic mixing changes the mass eigenstates and interactions of the vector bosons.  What follows is a brief synopsis of the results in \cite{Cassel:2009pu}, see also \cite{Feldman:2007wj}.  
After diagonalizing the kinetic terms and going to the mass eigenstate basis, the SM neutral current interactions are modified.
Absorbing the gauge coupling constants into the definition of the currents, the 
neutral  current interactions are
\begin{eqnarray}
\LL_{\text{int}}
 = V^\mu_{\text{Gauge}} J_\mu
  = V^\mu_{\text{Mass}}  \mathcal{M} J_\mu, 
\end{eqnarray}
where the notation for the gauge and mass eigenstates is
\bea
V^\mu_{\text{Gauge}}= \begin{pmatrix}
A^\mu \\ Z^{0\,\mu} \\ A'{}^\mu \end{pmatrix}
\qquad
V^\mu_{\text{Mass}}= \begin{pmatrix}
A^\mu \\ Z^\mu \\ Z'{}^\mu \end{pmatrix}
\end{eqnarray}
and the currents are
\begin{eqnarray}
J^\mu=
\begin{pmatrix}
e J^\mu_{\text{EM}} \\ g/c_{\text{w}} J^\mu_{Z^0} \\ \tilde{g}J^\mu_{A'} \end{pmatrix}
\end{eqnarray}
with the diagonalization matrix
\begin{eqnarray}
\mathcal{M} =  
\begin{pmatrix}
1 & 0 & 0 \\  -c_{\text{w}} t_\epsilon s_\xi & s_{\text{w}} t_\epsilon s_\xi + c_\xi &  s_\xi / c_\epsilon \\ -c_{\text{w}} t_\epsilon c_\xi & s_{\text{w}} t_\epsilon c_\xi - s_\xi  &  c_\xi / c_\epsilon
\end{pmatrix}.
\eea
$c, s, t$ stand for cosine, sine and tangent respectively, and $c_{\text{w}}$  and $s_{\text{w}}$ are the cosine and sine of the weak mixing angle.  The photon's interactions are  unaltered due to its residual gauge invariance. The angle $\xi$ is  defined as
\begin{eqnarray}
\tan2\xi &=& \frac{2 \Delta (m_{Z'}^2 - \mzt^2)}{(m_{Z'}^2 -\mzt^2)^2 - \Delta^2}, \\
\Delta &=& - \mzt^2 \sin\theta_{\text{w}} \tan\epsilon 
\end{eqnarray}
with
$\mzt=   m_{W^\pm}/ \cos \theta_{\text{w}}$.
After changing to the mass eigenstate basis, the physical mass of the $Z^0$, $m_Z$, is
\begin{eqnarray}
\label{Eq: ZMass}
m_Z^2 &=& \frac{\mzt^2  - m_{Z'}^2 \sin^2 \xi }{\cos ^2\xi}
\end{eqnarray}
and the physical mass of the new vector boson is
\begin{eqnarray}
m_{Z'}^2 &=& 
 m_{A'}^2 \frac{c^2_\xi}{c^2_\epsilon}+\mzt^2 s_\xi^2\left(1 + \frac{s_{\text{w}} t_\epsilon}{t_\xi} \right)^2 .
\end{eqnarray}

These corrections to the SM  neutral currents and to the mass of the $Z^0$ place model independent bounds on $( m_{Z'},\epsilon)$.    
The next section describes the SM measurements that are sensitive to these modified neutral current interactions.

\section{Precision SM Measurements}
\label{Sec: Experiments}

Virtual $Z'$ exchange modifies measured observables such as  
 Bhabha scattering, forward-backward asymmetry measurements, $m_Z$, and the total hadronic cross sections.   The mass of the $Z^0$ is the most powerful single measurement but the constraint is augmented by other measurements at and above the $Z^0$ pole.
 Additionally,  if the $Z'$ has a sizeable branching ratio back to the Standard Model, 	resonant production of the $Z'$ bounds  the parameter space at specific energies.
 
The strongest constraint on $\epsilon$ comes from the shift of the $Z^0$ mass \cite{Amsler:2008zzb}.  
Notice that  $\xi$ in Eq. \ref{Eq: ZMass}  changes sign as $m_{Z'}$ goes through $\mzt$, meaning that the corrections to the $Z^0$ mass vanish at this point.
Defining  \mbox{$\delta m =m_{Z'} - m_{Z^0}$,} the correction to the $Z^0$ mass is given by
\bea
\frac{m_Z - m_{Z^0}}{m_{Z^0}} = - t^2_\xi \frac{ \delta m}{m_{Z^0}} + \OO\left( \frac{\delta m^2}{m_{Z^0}^2}\right) \le 
2.5 \times 10^{-5}
\eea
so that as $\delta m \rightarrow 0$, there is no bound on $\epsilon$ resulting from the measurement of $m_{Z}$.   
There is a reduction in the limits on $\epsilon$ for
\begin{eqnarray}
m_{Z'} \simeq m_{Z^0}\pm  0.1 \GeV
\end{eqnarray}
where other measurements  must take over for the $Z^0$ mass measurement.


 $e^+e^-$ colliders measure the SM neutral current interactions and when $m_{Z'} \lsim \sqrt{s} \le \mzt $, the $Z'$ couples dominantly to the electromagnetic current, which causes a kink to appear in the running of the fine structure constant. 
 Differential Bhabha scattering measures $\alpha_{\text{EM}}(q^2)$ and there is a wealth of data from experiments such as OPAL \cite{Abbiendi:2003dh}, DELPHI \cite{Abdallah:2005ph}, SLD \cite{Pitts:1994he}, TASSO \cite{Braunschweig:1987py}, CELLO \cite{CELLO:1989rfa} and TRISTAN \cite{Arima:1996ev}.  
  As a result, the new vector boson changes the predictions for differential Bhabha scattering.  All the experiments above have a large forward bin of $\cos \theta \gsim 0.9$, so only a small range of $q^2$ is probed at each experiment.  
The forward bin normalizes the luminosity and cannot be used as a constraint, thereby limiting the power of these measurements.  Differential Bhabha scattering for $\sqrt{s} \le \mzt$ is not useful, but provides additional constraints at and above the $Z^0$ pole, where corrections to $m_Z$ are less powerful.

 In addition to differential Bhabha scattering, the forward-backward
asymmetries for the bottom, charm, muon and tau are measured at $m_{Z^0}$,  effectively fixing $\alpha_{\text{EM}}(\mzt)$ and $\sin^2 \theta_{\text{w}}$ \cite{:2003ih}.
The modification to the SM neutral currents alter $\alpha_{\text{EM}}(\mzt)$ and $\sin^2 \theta_{\text{w}}$ and  leads to a conflict with other SM predictions, most notably $\Gamma_{Z^0}$ and $\sigma_{\text{had}}\equiv \sigma(e^+ e^-\rightarrow \text{hadrons})$, {\it i.e.}  
\mbox{$\Gamma_{Z^0}=\Gamma_{Z^0}( \alpha_{\text{EM}}(\mzt), \sin^2 \theta_{\text{w}}, G_F)$.}


%

Resonant and on-shell production of the $Z'$ can be relevant even if there is a small width directly back to the SM.  
The $Z'$ has a decay width  into the SM and dark sectors given by
\begin{eqnarray}
\Gamma_{Z' \text{ SM}} \simeq \frac{\epsilon^2 g^2 m_{Z'}}{4\pi}\qquad \Gamma_{Z'\text{ dark}}\simeq \frac{\tilde{ g}^2 m_{Z'}}{4\pi} .
\end{eqnarray}
The width of the $Z'$ into the dark sector is unknown; however, given that bounds from the $Z^0$ mass set $\epsilon \lsim \OO(10^{-2})$ and $\tilde{g}$ is bounded by $\OO(1)$, there can be a detectable width for the $Z'$ back into the SM.
As a way to parameterize these effects, two different dark sector widths are used in setting limits
\begin{eqnarray}
 \Gamma_{Z'\text{ dark}}= \begin{cases} 10^{-2}\, m_{Z'}&  \text{ (wide) } \\
 0 & \text{ (narrow) }
 \end{cases} .
\end{eqnarray}
On-shell production of the $Z'$ is calculated using MadGraph 4.4.32 \cite{MadGraph}.
Only the interference between the $Z'$ and the SM is explicitly computed by zeroing out
the $|\text{SM}|^2$ and $|Z'|^2$ squared matrix elements.  This results in a  deviation from the SM that scales as $\epsilon^2$
and the calculations can be  compared with measurements using the methods described in the next section.

The total hadronic cross sections, $\sigma_{\text{had}}$,  are measured at LEP2 with $\mzt \le \sqrt{s}\le  207 \GeV$ \cite{Abbiendi:2003dh,Abdallah:2005ph} as well as at many other experiments with $22 \GeV \le \sqrt{s}\le  64 \GeV$ \cite{Janot:2004cy}.   
These measurements provide additional bounds because the results from differential Bhabha scattering are not reported at every energy.  While the error bars are large compared to the differential Bhabha scattering, resonant $Z'$ production enhances sensitivity if $m_{Z'}\simeq \sqrt{s}$.
Radiative return processes involving the $Z'$ could in principle constrain the theory for $\sqrt{s}$ away from $m_{Z'}$; however, these never provide competitive measurements.

The $Z^0$ can have exotic decays into the hidden sector and assuming that there are no mass thresholds in the hidden sector between $\mzt$ and $m_{Z'}$, then
\begin{eqnarray}
\Gamma_{Z^0 \text{ exotic}} \simeq  t_\xi^2 \frac{\mzt}{m_{Z'}} \Gamma_{Z'}  .
\end{eqnarray}
The $Z^0$ line shape measurement constrains $\Gamma_{Z^0}$ in a model independent manner giving a bound on $\Gamma_{Z'}$ \cite{Amsler:2008zzb}.
 The bound on the width of the $Z'$ is shown in Fig.~\ref{Fig:GammaBounds}.  

\begin{figure}[htb]
\begin{center}
 \includegraphics[width=3.in]{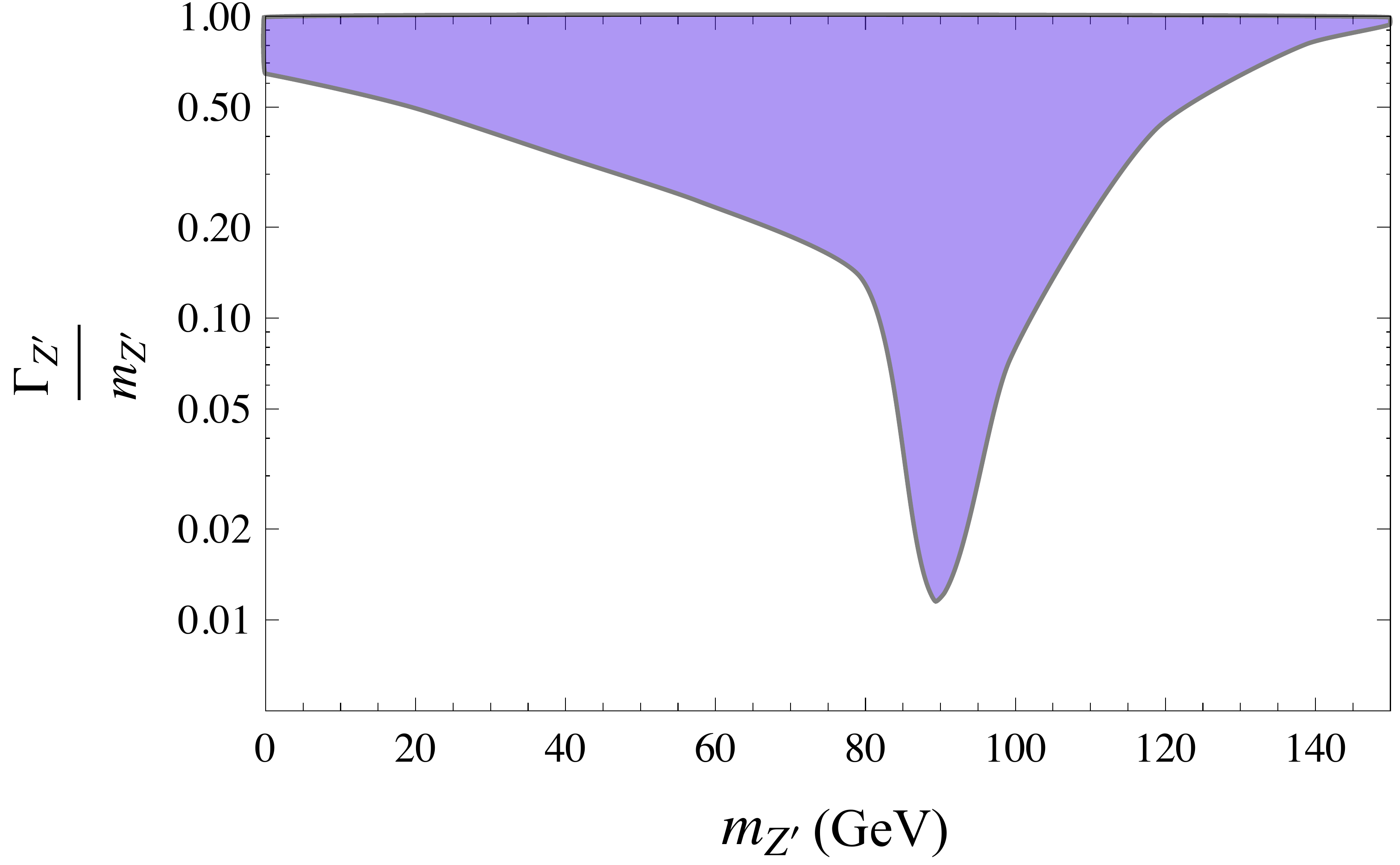}
 \caption{The model independent upper bounds on $\Gamma_{Z'}$ arising from the line shape of the $Z^0$. }
   \label{Fig:GammaBounds}
   \end{center}
 \end{figure}

In addition to precision $e^+e^-$ measurements,  direct searches at the Tevatron can produce on-shell $Z'$s.   This letter finds that the Tevatron's sensitivity is just beneath precision electroweak results even assuming  $\Gamma_{Z'\text{ dark}}=0$
\cite{CDF:2007sb, Cassel:2009pu}.

\section{Results}
\label{Sec: Results}

The calculated deviations from SM measurements detailed in the previous section can be combined into an exclusion plot for the $(m_{Z'},\epsilon )$ plane.
This letter uses tree-level predictions for the full theory, $X_{\text{th}}(i, \epsilon, m_{Z'})$, for a measurement, $i$,  and subtracts off the SM contribution, $X_{\text{th SM}}(i)$, from the calculation to compute the residual change from the SM prediction
\begin{eqnarray}
\delta_{\text{th}}(i, \epsilon, m_{Z'}) = \frac{X_{\text{th}}(i, \epsilon, m_{Z'})-X_{\text{th SM}}(i)}{X_{\text{th SM}}(i)}.
\end{eqnarray}
The theory residuals are compared against  the experimental residuals defined
as 
\begin{eqnarray}
\delta_{\text{exp}}(i)= \frac{X_{\text{exp}}(i)-X_{\text{exp SM}}(i)}{X_{\text{exp SM}}(i)}\quad
\end{eqnarray}
where  $X_{\text{exp SM}}(i)$ is the SM prediction presented for each result.  $X_{\text{exp SM}}(i)$ typically includes higher order corrections to the SM prediction and detector-dependent corrections; therefore, the comparison of the theory and experiment respective residuals can be made reliably.

The regions in the $(m_{Z'},\epsilon)$ parameter space consistent with precision SM measurements   are found by performing a global fit to the SM parameters.
This letter uses a $\chi^2$ test that is a function of $\epsilon$, $m_{Z'}$ and the Standard Model parameters, $\alpha_{\text{EM}}(\mzt)$, $G_F$ and $\sin^2\theta_{\text{w}}$.  $G_F$ is fit by $\mu$ decay and does not vary in practice. 
The global $\chi^2$ is
\bea
\!\chi^2(\epsilon, m_{Z'}; \text{SM Param}) \!= \!\sum_i \!\frac{\left(\delta_{\text{th}}(i, \epsilon, m_{Z'}) - \delta_{\text{exp}}(i) \right)^2}{\left(\hat{\sigma}_{\text{exp}}(i) \right)^2}
\eea
with the accuracy of each measurement being
\begin{eqnarray}
\hat{\sigma}_{\text{exp}}(i) = \frac{\sigma_{\text{exp}}(i)}{X_{\text{exp SM}}(i)} .
\end{eqnarray}
In order to not dilute the $\chi^2$ by superfluous measurements that have no {\it a priori} possibility of constraining a theory with a given $(m_{Z'},\epsilon )$, 
only experiments that had an expected significance, $E(i; \epsilon, m_{Z'})$, satisfying
\bea
E(i; \epsilon, m_{Z'})= \frac{|\delta_{\text{th}}(\epsilon, m_{Z'}) |}{\hat{\sigma}_{\text{exp}}} >  E_{\text{min}}
\eea
are included in the $\chi^2$ with $E_{\text{min}}=0.5$.   The results in this letter are insensitive to the exact choice of $E_{\text{min}}$ and this value was chosen for convenience.

For $m_{Z'} \gg 200 \GeV$, the effects of the new vector boson can be encapsulated in terms of local operators and coincide with the precision electroweak analyses, {\it e.g. } the $S$, $T$ parameters \cite{Peskin:1991sw,Altarelli:1990zd,:2005ema} or more recently \cite{Barbieri:2004qk,Han:2004az}. 
For $m_{Z'} \lsim m_{Z^0}$, the bounds are close to those from \cite{Feldman:2007wj} which only use the constraint from $m_{Z}$. 
  
Fig. \ref{Fig:AprimeBounds} shows the 95\% confidence level (CL) excluded regions in the $(m_{Z'},\epsilon)$ plane obtained in this study.   Wide $Z'$s are best constrained by the mass of the $Z^0$ for most of the parameter space.  The exception occurs near the $Z^0$ mass.  The  forward-backward asymmetries, $\Gamma_{Z^0}$, and $\sigma_{\text{had}}$ augment the limits when the corrections to the $Z^0$ mass vanish and also for $m_{Z'}\simeq 200 \GeV$ where LEP2 forward-backward measurements are more constraining than the $Z^0$ mass.
Limits on narrow $Z'$s are enhanced for $m_{Z'}\simeq \sqrt{s}$ for the numerous $e^+e^-$ experiments.  The forward-backward asymmetries, hadronic cross section and differential Bhabha scattering measurements provide the  additional constraints.
The peaks appearing in the exclusion region can be traced to experimental energies at which various experiments were conducted.
The constraint on $m_{Z'}$ near the $Z^0$ is illustrated in the inset of Fig.~\ref{Fig:AprimeBounds} .
For comparison, the bounds from $(g-2)_\mu$, and model-dependent  $e^+e^- \rightarrow \gamma  Z'\rightarrow \gamma \mu^+ \mu^-$ BaBar searches from  \cite{BaBar:2009,Essig:2010xa} are shown.  

\begin{figure}[t]
\begin{center}
 \includegraphics[width=3.5in]{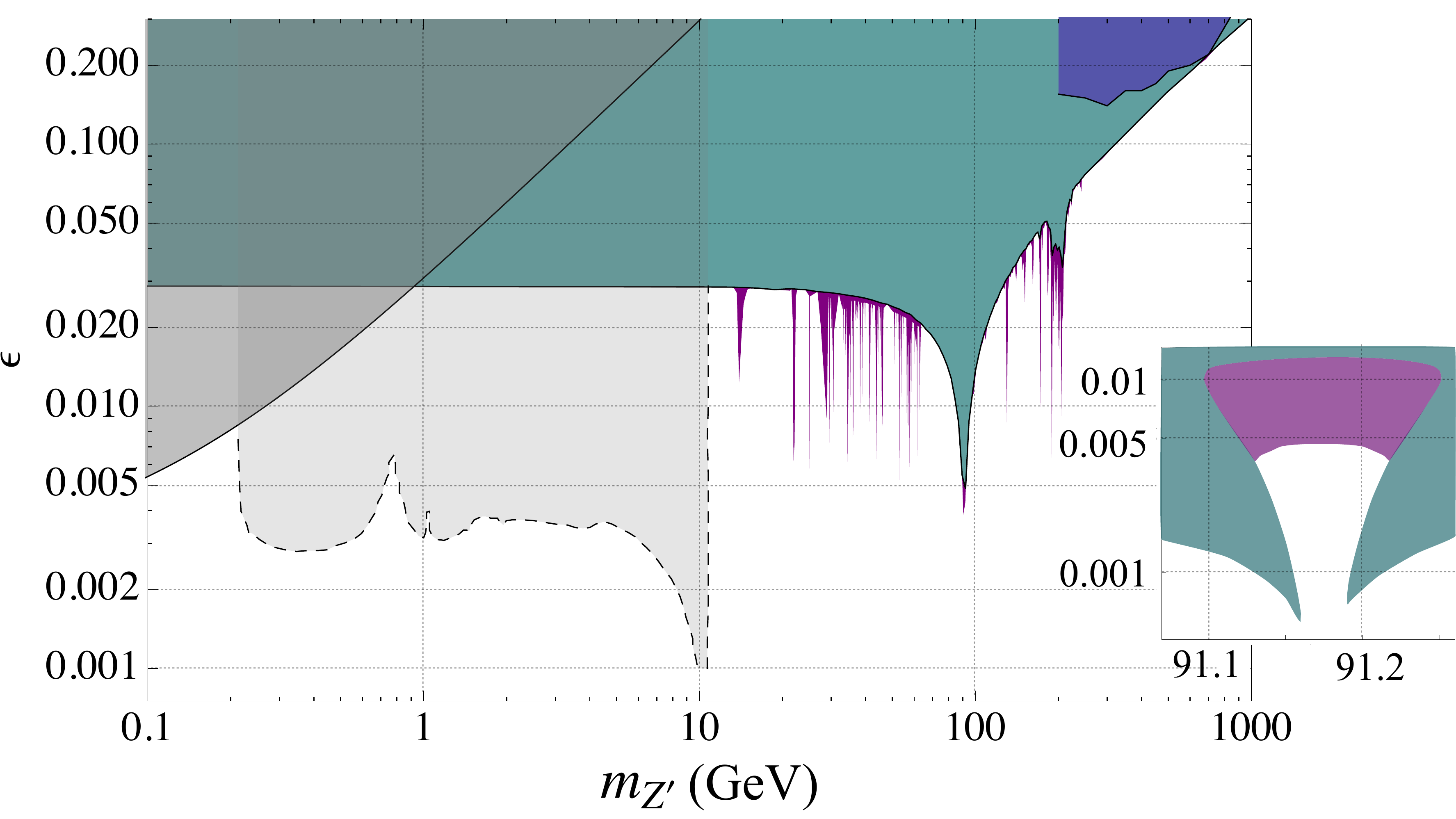}
 \caption{95\% CL exclusions in the $(m_{Z'},\epsilon)$. The cyan region is excluded for a ``wide'' $Z'$ and the purple region is for a ``narrow'' $Z'$.  The blue region shows the bounds placed by CDF on direct production of $Z'$s. The inset illustrates the constraints on $m_{Z'}$ near the $Z^0$ pole. 
The bound from the $(g-2)_\mu$ is shown in dark grey and the light grey, dashed region shows the sensitivity from model dependent BaBar searches. 
 }
   \label{Fig:AprimeBounds}
   \end{center}
 \end{figure}

The model-independent limits on kinetic mixing were computed in this letter and found to be $\epsilon\lsim 0.03$ for most of the mass range studied, $1 \GeV < m_{Z'} < 200 \GeV$. The possible use of radiated return to place tighter constraints on $Z'$ was investigated at both LEP1 and LEP2 energies, however this channel did not help place tighter bounds on kinetic mixing.  Even with the constraints found in this letter, there still is a vast parameter space available for a kinetically mixed vector boson to mediate
interactions between a dark sector and the SM.    The current program of searching for model dependent decay modes at low energy experiments will augment these model independent limits for $m_{Z'} \lsim 10 \GeV$.  For higher energies, only the LHC  will provide additional information for $200 \GeV \lsim m_{Z'} \lsim 3 \TeV$ \cite{Cassel:2009pu}.   The relatively weak limits for $m_{Z'} \gsim 10\GeV$ motivates new high intensity $e^+e^-$ experiments to potentially discover new interactions of this form.

\section*{Acknowledgements}
We would like to thank D. E. Kaplan, M. Lisanti,  and M. Peskin for numerous useful conversations.  AH, EI, and  JGW are supported by the US DOE under contract number DE-AC02-76SF00515.  
AH, EI, and JGW receive partial support from the Stanford Institute for Theoretical Physics.  
JGW is partially supported by the US DOE's Outstanding Junior Investigator Award.


\end{document}